\documentclass[11pt,a4paper,portrait]{extarticle}
\usepackage{amsmath}
\usepackage{authblk}

\title{An Intuitive Procedure for Converting PDA to CFG, by Construction of Single State PDA}
\author{
Arjun Bhardwaj\thanks{School of Computing and Electrical Engineering, Indian Institute of Technology Mandi, India, bhardwaj.arjun.14@gmail.com} \:\: N.S. Narayanaswamy\thanks{Associate Professor, Department of Computer Science and Engineering, Indian Institute of Technology Madras, India, swamy@iitm.ac.in}
}
\date{Novemeber 1, 2013}

\begin{document}

\maketitle

\section{Introduction}
We present here the proof for an alternative procedure to convert a Push Down Automata (PDA) into a Context Free Grammar (CFG). The procedure involves intermediate conversion to a single state PDA. In view of the authors, this conversion is conceptually simpler than the approach presented in \cite{ull} and can serve as a teaching aid for the relevant topics. For details on CFG and PDA, the reader is referred to \cite{ull}.

\section{Construction Procedure : Multistate PDA to Single State PDA}

For a Push Down Automata (PDA) that accepts by null store : N = $(Q,\Sigma,\Gamma,\delta,q_o,Z_o)$, we can construct an equivalent single state PDA : $N_s$ = $(Q_s,\Sigma,\Gamma_s,\delta_s,q_m,Z_s)$,
such that $Q_s = \{q_m\}$, and $\Gamma_s$ is of the form [pXq] where $p,q \in Q$, $X \in\Gamma$. 
$\delta_s$ is constructed by following rules :\\
\begin{enumerate}
\item For every $\delta(p,a,X)=(q,\epsilon)$, we add $\delta_s(q_m,a,[pXq]) = (q_m,\epsilon)$
\item For every $\delta(p,a,X)=(q,B_1,B_2...B_l)$, $B_i \in \Gamma$, we add $\delta_s(q_m,a,[pXq])=(q_m,[qB_1s_1][s_1B_2s_2]...[s_lB_lq])$, $s_i \in Q$
\item $\delta(q_m,\epsilon,Z_s)=(q_m,[q_oZ_os_i])$ for all $s_i \in Q$
\end{enumerate}

\subsection {Proof}

We provide a formal proof for the following claim : $L(N) = L(N_S)$

\subsubsection{(IF part) $L(N) \subseteq L(N_s)$}

To prove this, we first prove the following claim :\\
$if : (p,w,X) \overset{*}{\mapsto} (q,\epsilon,\epsilon)$\\
$then : (q_m,w,[pXq]) \overset{*}{\mapsto} (q_m,\epsilon,\epsilon)$
This is proved by induction on the number of steps in which X is popped off by the one state automata.\\

Base : $(p,a,X) \overset{1}{\mapsto} (q,\epsilon,\epsilon)$. This is possible only if $a \in (\Sigma \cup {\epsilon})$ and $\delta(p,a,X)\in(q,\epsilon)$. 
Then, according to construction, there is a corresponding entry in transition table of multistate automaton i.e. $\delta(q_m,a,[pXq])\in(q_m,\epsilon)$. \\

Inductive step : assume that if $(p,w,X) \overset{j}{\mapsto} (q,\epsilon,\epsilon)$, then $(q_m,w,[pXq]) \overset{*}{\mapsto} (q_m,\epsilon,\epsilon)$; for $j \leq n$. Consider $a\in \Sigma\cup{\epsilon}$ and $\beta \in \Sigma^*$ such that $(p,a\beta,X) \overset{1}{\mapsto} (t,\beta,B_1...B_k) \overset{n}{\mapsto} (q,\epsilon,\epsilon)$ i.e. X is popped off in n+1 steps.
Since $\beta$ is the string that gets processed as PDA N pops off all of $B_i$, we can decompose $\beta$ into $\beta_1\beta_2...\beta_k$ such that $\delta(t,\beta_1,B_1) \overset{n_1}{\mapsto} (t_1,\epsilon)$, $\delta(t_{i-1},\beta_i,B_i) \overset{n_i}{\mapsto} (t_i,\epsilon)$ ... $\delta(t_{k-1},\beta_k,B_k) \overset{n_i}{\mapsto} (q,\epsilon)$, where $\sum_{i=1}^{k}n_k = n$.\\
The rule for single state automaton corresponding to one responsible for $(p,a\beta,X) \overset{1}{\mapsto} (t,\beta,B_1...B_k)$, yields $(q_m,a\beta,X) \overset{1}{\mapsto} (q_m,\beta, [tB_1t_1]... [t_{k-1}B_kq])$. further, by induction hypothesis, $(q_m,\beta_1,[tB_1t_1]) \overset{*}{\mapsto} (q_m,\epsilon,\epsilon)$, $(q_m,\beta_i,[t_{i-1}Bt_i]) \overset{*}{\mapsto} (q_m,\epsilon,\epsilon)$ and $(q_m,\beta_k,[t_{k-1}B_kq])$. thus, $(q_m,\beta_1...\beta_k,[pB_1t_1]...[t_k-1B_kq]) \overset{*}{\mapsto} (q_m,\epsilon,\epsilon)$, proving the claim.\\
As a special case of the claim, consider : $(p,w,Z_o) \overset{*}{\mapsto} (q,\epsilon,\epsilon)$ i.e. $w \in L(N)$. then for $N_s$, $(q_m,w,[pZ_oq]) \overset{*}{\mapsto} (q_m,\epsilon,\epsilon)$. 
Using construction rule 3, $(q_m,w,Z_s) \overset{*}{\mapsto} (q_m,\epsilon,\epsilon)$ i.e. $w \in L(N_s)$

\subsubsection{(ELSE part) $L(N_s) \subseteq L(N)$}
To prove this, we first prove the following claim :\\
$if : (q_m,w,[pXq]) \overset{n}{\mapsto} (q_m,\epsilon,\epsilon)$\\
$then : (p,w,X) \overset{*}{\mapsto} (q,\epsilon,\epsilon)$\\
We provide a proof by induction on the no of steps taken by single state automata to pop off [pXq].\\

Base : if $(q_m,w,[pXq]) \overset{1}{\mapsto} (q_m,\epsilon,\epsilon)$. Then there must be a rule in $N_s$ : $\delta_s(q_m,w,[pXq])=(q_m,\epsilon)$, where $w \in \Sigma\cup{\epsilon}$. 
According to construction, this rule derives from $\delta(p,w,X)=(q,\epsilon)$ in N. thus, $(p,w,X)\overset{1}{\mapsto} (q,\epsilon,\epsilon)$.\\

Inductive case : assume that if : $(q_m,w,[pXq]) \overset{j}{\mapsto} (q_m,\epsilon,\epsilon)$ then : $(p,w,X) \overset{*}{\mapsto} (q,\epsilon,\epsilon)$, for $j \leq n$. Consider $(q_m,a\beta,[pXq])$ $\overset{1}{\mapsto}$ $(q_m,\beta,[s_1B_1s_2][s_2B_2s_3]...[s_kB_kq])$ $\overset{n}{\mapsto}$ $ (q_m,\epsilon)$. 
Since all the symbols stacked after the first move in place of $[pXq]$ will all be popped off, we can decompose $\beta = \beta_1...\beta_k$ such that $(q_m,\beta_i,[s_iB_is_{i+1}]) \overset{n_i}{\mapsto} (q_m,\epsilon,\epsilon)$, where $\sum_{i=1}^{k}n_i = n,$ $s_{k+1} = q$. by inductive hypothesis, each of such transitions corresponds to $(s_i,\beta_i,B_i) \overset{*}{\mapsto} (s_{i+1},\epsilon,\epsilon)$ for PDA N. 
Also, the first transition of $N_s$ corresponds to presence of rule $\delta(p,a,X) = (s1,B_1B_2...B_k)$. thus, $(p,a\beta,X) \overset{1}{\mapsto} (s_1,\beta_1...\beta_k,B_1B_2...B_k) \overset{*}{\mapsto} (s_2,\beta_2...\beta_k,B_2...B_k) ... \overset{*}{\mapsto} (s_k,\beta_k,B_k) \overset{*}{\mapsto} (q,\epsilon,\epsilon)$. 
Thus $a\beta \in L(N)$, completing the proof of the above mentioned claim.\\\\
Consider $w \in L(N_s)$. Then for some $s_1\in Q$, $(q_m,w,Z_s) \overset{1}{\mapsto} (q_m,w,[q_oZ_os_1]) \overset{*}{\mapsto} (q_m,\epsilon,\epsilon)$. Using the claim, we can assert $(q_o,w,Z_o) \overset{*}{\mapsto} (s_1,\epsilon,\epsilon)$. \\

Thus, for a single state PDA $N_s$ constructed from multistate PDA N by the procedure described above, $L(N) = L(N_s)$.

\section{construction : single state PDA to grammar G}

For a single State PDA $N_s$ = $(Q_s,\Sigma,\Gamma_s,\delta_s,q_m,Z_s)$, we construct a grammar G = (V,T,P,S), such that L(G) = L($N_s$), where V = $\Gamma_s$, T = $\Sigma$ and S = $Z_s$. the following rules outline the set of productions, P :\\
\begin{enumerate}
\item For every $(q_m,a,Z)=(q_m,\gamma)$, we add $Z \rightarrow a\gamma$, where $a\in\Sigma\cup\{\epsilon\}$, $Z\in\Gamma$, $\gamma\in\Gamma_s^*$
\end{enumerate}

\subsection{proof}
Here we describe the proof for the following claim : $L(G) = L(N_s)$

\subsubsection{(IF part) $L(G) \subseteq L(N_s)$}
To prove this we first prove the following claim :\\
$if : A \overset{*}{\rightarrow} w$\\
$then : (q_m,w,A) \overset{*}{\mapsto} (q_m,\epsilon,\epsilon)$, \\
where $A\in V, w \in \Sigma^*$. Proof is by induction on the number of steps in the derivation of w.\\

Base : $A \overset{1}{\rightarrow} w$. 
This indicates the presence of identical production in P. By construction, we can conclude the presence of $(q_m,w,A)=(q_m,\epsilon)$. 
Thus, $(q_m,w,A) \overset{1}{\mapsto} (q_m,\epsilon,\epsilon)$.\\

Induction : assume that if $A \overset{j}{\rightarrow} w$, then $(q_m,w,A) \overset{*}{\mapsto} (q_m,\epsilon,\epsilon)$, for $j \leq n$. 
Consider the left derivation of the string $v\in\Sigma^*$, such that $A \overset{1}{\rightarrow} aY_1Y_2...Y_k \overset{n}{\rightarrow} v$, where $Y_i \in V$, $a \in \Sigma\cup\{\epsilon\}$. 
Since v has a leftmost derivation, each $Y_i$ is replaced by a part of the terminal string $v$ i.e. $Y_i \overset{n_i}{\rightarrow} z_i$, where $az_1z_2...z_k=v$, $\sum_{i=1}^{k}n_k=n$. by inductive hypothesis, $(q_m,z_i,Y_i) \overset{*}{\mapsto} (q_m,\epsilon,\epsilon)$. 
Further the first step in derivation indicates the presence of rule $\delta_s(q_m,a,A) = (q_m,Y_1Y_2...Y_k)$ in $N_s$, such that $(q_m,a,A) \overset{q}{\mapsto} (q_m,\epsilon,Y_1Y_2...Y_k)$. 
Thus, $(q_m,A,az_1...z_k) \overset{1}{\mapsto} (q_m,Y_1...Y_k,z_1...z_k) \overset{*}{\mapsto} (q_m,Y_2...Y_k,z_2...z_k) ... \overset{*}{\mapsto} (q_m,\epsilon,\epsilon)$\\
For the special case of $A=Z_s$, the claim yields : if $Z_s \overset{*}{\rightarrow} w$, then $(q_m,w,Z_s) \overset{*}{\mapsto} (q_m,\epsilon,\epsilon)$. 
Thus, if w is derived by G, then it is accepted by $N_s$ i.e. $L(G) \subseteq L(N_s)$.

\subsubsection{(THEN part) $L(N_s) \subseteq L(G)$}
To prove this we first prove the following claim :\\
$if : (q_m,w,A) \overset{*}{\mapsto} (q_m,\epsilon,\epsilon)$
$then : A \overset{*}{\rightarrow} w$, where $A\in V, w \in \Sigma^*$\\
This is proved by induction on the number of state transitions taken by $N_s$ to reach $(q_m,\epsilon,\epsilon)$.\\

Base : $(q_m,w,A) \overset{1}{\mapsto} (q_m,\epsilon,\epsilon)$. This is possible only due to the presence of rule $\delta(q_m,w,A)=(q_m,\epsilon)$. According to construction, G has the following production : A$\rightarrow$w. Thus, w is derived by G.

Induction : Assume that if $(q_m,w,A) \overset{j}{\mapsto} (q_m,\epsilon,\epsilon)$, then $A \overset{*}{\rightarrow} w$, for $j \leq n$. Consider $a\in\Sigma\cup\{\epsilon\}$ and $v\in\Sigma^*$, such that $(q_m,av,A) \overset{1}{\mapsto} (q_m,v,A_1A_2...A_k) \overset{n}{\mapsto} (q_m,\epsilon,\epsilon)$.
The fist step indicated the presence of the rule $\delta(q_m,a,A)=(q_m,A_1A_2...A_k)$ in $N_s$ and correspondingly the production $A \rightarrow aA_1A_2...A_k$ in P.
Since all of the stack symbols $A_i$ are eventually popped off, we can decompose $v$ into $v_1v_2...v_k$ such that $(q_m,v_i,A_i)\overset{n_i}{\mapsto}(q_m,\epsilon,\epsilon)$, where $\sum_{i=1}^{k} n_i = n$ and $v_i \in \Sigma^*$.
By inductive hypothesis, this indicates the presence of derivations $A_i \rightarrow v_i$. 
Thus, we have a sequence of derivations $A \overset{1}{\rightarrow} aA_1...A_k \overset{*}{\rightarrow} av_1A_2...A_k \overset{*}{\rightarrow} av_1...v_k$. Thus proved.

Consider as a special case A=$Z_s$. Then, if $av$ is accepted by $N_s$, it is derived by G. We can conclude $L(N_s) \subseteq L(G)$.\\

So, $L(N_s) = L(G)$.

\end{document}